\documentclass[sigconf,nonacm]{acmart}

\usepackage{graphicx}
\usepackage{balance}

\renewcommand\footnotetextcopyrightpermission[1]{}
\setcopyright{none}

\title{When Visibility Outpaces Verification: Delayed Verification and Narrative Lock-in in Agentic AI Discourse}

\author{Hanjing Shi}
\email{hasa23@lehigh.edu}
\affiliation{
  \institution{Lehigh University}
  \city{Bethlehem}
  \state{Pennsylvania}
  \country{USA}
}

\author{Dominic DiFranzo}
\email{djd219@lehigh.edu}
\affiliation{
  \institution{Lehigh University}
  \city{Bethlehem}
  \state{Pennsylvania}
  \country{USA}
}

\begin{document}

\begin{abstract}
Agentic AI systems—capable of autonomous planning and execution—pose distinctive challenges for trust calibration. Long before direct interaction, users' expectations are shaped by public discourse on social platforms. However, platform-visible engagement signals (e.g., upvotes) may inadvertently create a "credibility proxy" that hinders critical evaluation. 

In this paper, we investigate the relationship between social proof and verification timing in online discussions of agentic AI. We analyze Reddit data from two distinct communities, operationalizing verification cues through reproducible lexical rules and modeling "time-to-first-verification" using right-censoring techniques. 

Our findings reveal a systemic "Popularity Paradox": discussions with high visibility experience significantly delayed or entirely absent verification cues compared to low-visibility threads. This lag creates a window of "Narrative Lock-in," where early, unverified claims crystallize into collective cognitive biases before evidence-seeking behaviors emerge. We discuss the implications of this "credibility-by-visibility" effect for AI safety and propose design interventions to introduce "epistemic friction" in engagement-driven platforms.
\end{abstract}

\maketitle

\section{Introduction}

Agentic AI systems---artificial intelligence systems that autonomously plan, decide, and execute multi-step tasks---are rapidly moving from research prototypes toward public-facing ecosystems \cite{He2025Plan, Balic2025Will}. Alongside this shift, public attention and concern are increasingly visible, including discourse about autonomy, safety, and responsibility \cite{Heaton2024ChatGPT, Hunter2026Bots}. Unlike earlier interactive systems, agentic AI is explicitly designed to act on users' behalf with limited oversight. As a result, questions of trust, verification, and responsibility become central: when, why, and on what basis should such systems be relied upon \cite{Lee2004Trust, He2025Plan}?

Crucially, trust in agentic AI often forms before direct interaction occurs. For many potential users, expectations about system capability, safety, and reliability are shaped through mediated exposure---news, social media posts, and online discussion forums \cite{Heaton2024ChatGPT, Naing2024Public, Xu2024Public}. Public discourse thus functions as a pre-interaction layer of trust calibration, where early impressions can guide later reliance even as direct experience becomes available \cite{Lee2004Trust}. This is especially consequential for emerging, weakly understood systems: early narratives can harden into community ``common sense'' before robust evidence is available.

In this paper, we examine how public discussions of agentic AI on Reddit evolve under the influence of platform-visible engagement signals such as post score and comment activity. We focus on evidence-seeking behavior---explicit requests for sources, links, or verification---and ask whether and when such behavior appears in highly visible discussions. Prior work shows that visible popularity shapes attention and evaluation, producing social influence bias and increasing vulnerability to unverified information \cite{Muchnik2013Social, Avram2020Exposure}. Building on this, we shift the question from \emph{whether} verification happens to \emph{when} it happens in the lifecycle of a discussion.

\textbf{Research questions.} We ask:
\begin{itemize}
  \item \emph{(RQ1) When a thread about agentic AI becomes highly visible, does evidence-seeking occur earlier in the thread lifecycle, or does it arrive later (or not at all) after discussion has already accumulated?}
  \item \emph{(RQ2) Does this visibility--verification relationship vary across communities with different discourse norms (exploratory vs.\ operational)?}
\end{itemize}

Rather than measuring individual trust directly, we study the informational conditions under which trust is formed. We operationalize verification timing as a structural property of discussion threads and analyze how it varies with platform visibility. We argue that delayed or absent verification in high-engagement threads is consistent with a form of collective cognitive bias: under novelty and epistemic uncertainty, groups substitute social proof for substantiation, implicitly delegating verification to the crowd \cite{Lee2004Trust, Muchnik2013Social}.

This paper contributes (1) a minimal, fully reproducible operationalization of thread-level verification cues and time-to-first-verification in public discourse; (2) an in-the-wild analysis showing that high visibility is associated with delayed or absent verification, characterizing a pre-interaction trust environment where credibility can be inferred from engagement before evidence is introduced; and (3) an account of how discourse context modulates this pattern, highlighting the role of community norms in shaping epistemic behavior around emerging AI.

\begin{figure*}[t]
  \centering
  \includegraphics[width=0.85\textwidth]{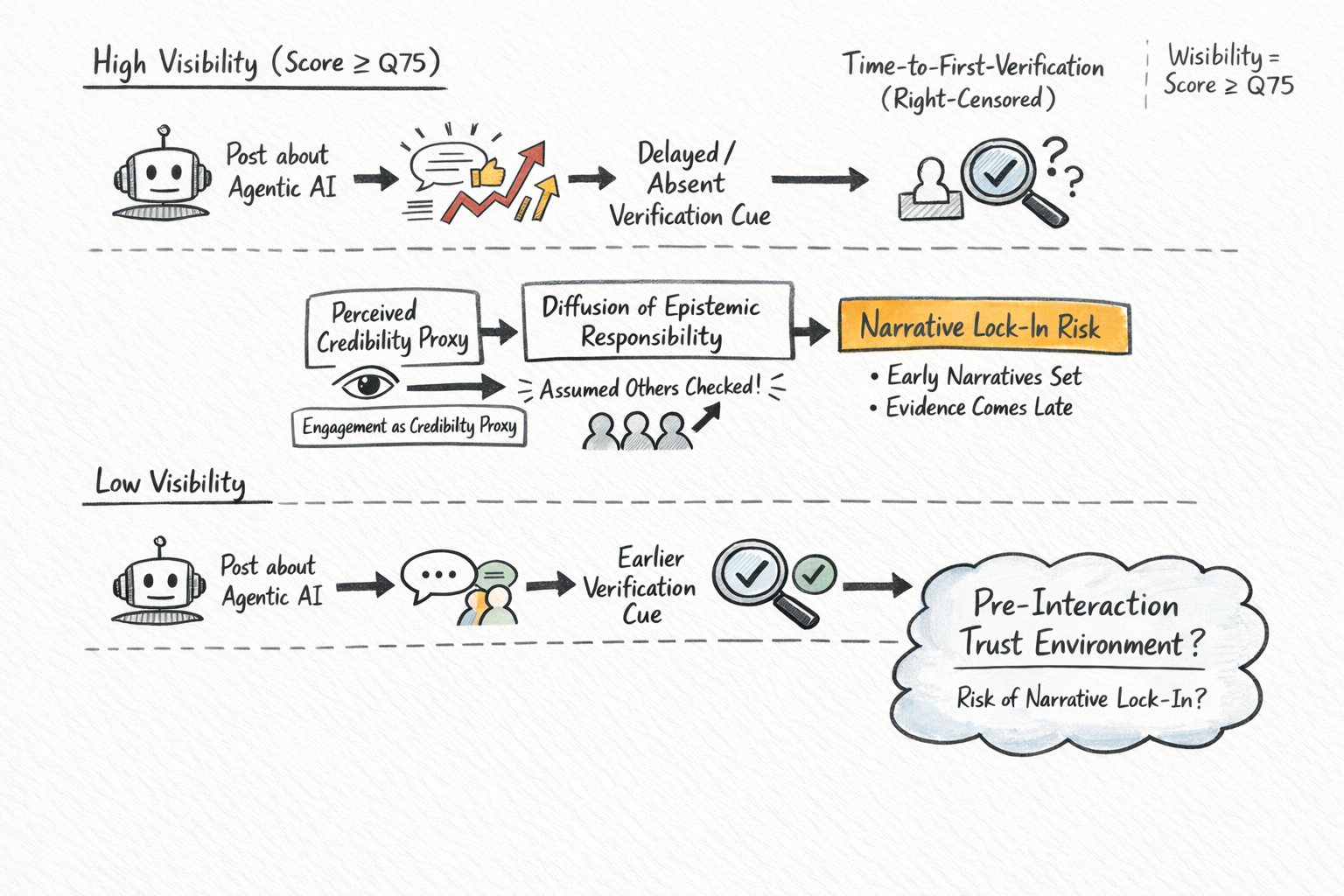}
  \caption{\textbf{Conceptual mechanism (not a quantitative result).} We study how platform-visible engagement (operationalized by score $\ge Q_{0.75}$ vs.\ lower) relates to the \emph{time-to-first-verification cue} in Reddit threads about agentic AI. Verification cues are identified using a high-precision lexical rule set; threads with no cue during the observation window are treated as right-censored. The figure illustrates a plausible pathway consistent with our findings: higher visibility can encourage credibility-by-visibility as a proxy, which can delay or suppress early evidence-seeking, creating a window where early narratives stabilize before substantiation appears.}
  \label{fig:mechanism}
\end{figure*}

\section{Related Work}

Our investigation intersects several established streams of research across human-computer interaction, cognitive science, and social computing. We situate our findings within the literature on trust in AI, the dynamics of misinformation and social proof, the role of community norms in shaping online behavior, and the use of design friction. Across these areas, prior work has largely examined \emph{what} people believe or \emph{how} they evaluate information; we focus on a complementary axis: \emph{when} evidence-seeking appears (or fails to appear) as public discourse unfolds, and how this timing differs under platform-visible engagement.

\subsection{Public Discourse and Trust in Agentic AI}
Trust is a fundamental determinant of human reliance on automated systems \cite{Lee2004Trust}. However, the emergence of agentic AI introduces new challenges for trust calibration, as these systems autonomously plan and execute multi-step tasks \cite{He2025Plan}. Recent scholarship emphasizes that trust in such systems is strongly influenced by conversational styles \cite{Belosevic2024Calibrating} and the perceived autonomy of the agents, which often exhibit complex collective behaviors \cite{Chen2025Unveiling}. Crucially, public perceptions and expectations of AI are rapidly negotiated in online forums and social media platforms long before widespread direct adoption \cite{Heaton2024ChatGPT, Balic2025Will, Hunter2026Bots}. Analyzing large-scale Reddit discussions reveals shifting public attitudes toward LLMs, moving from novelty to functional evaluation \cite{Naing2024Public, Xu2024Public}. 

Our work builds on this line of research by treating public discourse as part of a \emph{pre-interaction trust environment}: a setting where participants often lack direct access to ground truth and where credibility judgments may be made before hands-on experience is possible \cite{Lee2004Trust}. Rather than measuring trust attitudes directly, we operationalize a structural precursor to trust calibration---the timing and presence of explicit evidence-seeking---to characterize how early expectations can stabilize in agentic AI communities.

\subsection{Social Proof, Verification, and the Timing of Corrections}
When encountering novel information, users frequently rely on social heuristics. The presence of platform-visible engagement metrics serves as a powerful form of social proof, significantly increasing vulnerability to unverified information \cite{Avram2020Exposure, Muchnik2013Social}. The rapid dissemination of unverified claims---often manifesting as rumor cascades---can sometimes be mitigated when visible verification links are injected into the discourse \cite{Friggeri2014Rumor}. Yet, cognitive science demonstrates that the timing of such interventions is critical. Misinformation often exerts a continued influence even after credible corrections are issued \cite{Lewandowsky2012Misinformation}. Consequently, prebunking and accuracy nudges are generally more effective at improving truth discernment than post-hoc debunking \cite{Pennycook2021Shifting, Brashier2021Timing}. 

We connect these findings to public discourse about agentic AI by focusing on the \emph{temporal organization} of evidence-seeking. Prior work often evaluates whether corrections exist or whether people accept them; we ask whether verification appears early enough to shape the discussion trajectory in the first place. Methodologically, we treat threads without verification as a substantively important outcome rather than a missing case, motivating a right-censoring formulation that preserves the prevalence of ``never-verified'' discourse when analyzing time-to-first-verification.

\subsection{Community Norms, Moderation, and Vulnerability}
The verification dynamics we study are embedded in socio-technical environments where norms shape what counts as appropriate participation and accountability \cite{Dym2020, Fiesler2018, Nissenbaum2011}. Prior work shows that these norms and expectations strongly condition how communities respond to vulnerability, privacy, and sensitive disclosures \cite{Vitak2015, Wisniewski2018, Andalibi2016}. Some communities also develop norms and practices to support psychologically safe participation for members with constrained resources \cite{Israni2025, Andalibi2018, Malki2024}. Separately, platform policies and algorithmic classification systems can impose rigid categories that disproportionately burden marginalized identities, creating additional constraints on who is believed and what counts as legitimate evidence \cite{Haimson2015, Scheuerman2019, DeVito2021}. Communities therefore rely on moderation practices and governance architectures to sustain safety and legitimacy \cite{Jhaver2019, Seering2020, Chandrasekharan2017}. 

We extend this line of work by focusing on \emph{epistemic norms}: how discourse environments support or fail to support timely evidence-seeking in high-visibility threads about emerging AI. This emphasis matters for agentic AI because capability claims are often difficult to verify through shared benchmarks or reproducible demonstrations \cite{He2025Plan}. In such settings, community norms and platform affordances jointly shape whether evidence is demanded early, deferred until later, or never requested at all.

\subsection{Design Friction as an Epistemic Intervention}
To counteract the adverse effects of delayed verification, we draw upon the concept of design friction. In online safety and qualitative AI research, introducing deliberate friction has been shown to mitigate harm, reduce trauma, and improve the human decision-making process by shifting users from automatic to reflective cognitive states \cite{Ribeiro2023, Tseng2025, Distler2020}. We propose that integrating similar frictional elements into platform architectures could prompt earlier evidence-seeking behaviors, effectively disrupting collective reliance on engagement as a proxy for credibility. 

This framing treats verification timing as a design-relevant outcome: if platforms can shift when evidence is requested (and whether it appears at all), they may alter the pre-interaction trust environment in which agentic AI narratives form, without requiring platforms to adjudicate truth at the point of posting.

\section{Methods}

\subsection{Data Sources and Collection}

We study public Reddit discussions related to \emph{agentic AI}---AI systems described as capable of autonomously planning and executing tasks---using two subreddits with contrasting discourse norms: \texttt{r/moltbook} and \texttt{r/openclaw}. These communities were selected through exploratory inspection because both focus explicitly on agentic or autonomous AI systems, yet differ substantially in how such systems are discussed.

\texttt{r/moltbook} primarily hosts exploratory and interpretive discussions, including speculation about agent capabilities, personal experiences, and broader questions of autonomy and trust. In contrast, \texttt{r/openclaw} emphasizes operational and implementation-oriented discourse, such as deployment details, tooling, debugging, and verification practices. This contrast allows us to examine whether verification timing under platform-visible engagement varies across discourse contexts while holding the topic domain constant.

Data were collected programmatically using the Reddit API over a fixed observation window from January~1 to February~6, 2026. For each subreddit, we retrieved all posts created during this window and all comments associated with each post at the time of collection (i.e., the full comment tree available via the API snapshot). We treat each post together with its comment tree as a discussion thread.

Posts or comments whose bodies were unavailable (e.g., \texttt{[deleted]} or \texttt{[removed]}) were excluded from cue matching and timing. Such unavailable items may still be reflected in raw API-level comment counters, but they are not part of the retained text corpus analyzed here. Comments were de-duplicated. If a comment identifier was available, we de-duplicated by \texttt{(post\_id, comment\_id)}; otherwise we used the deterministic fallback key \texttt{(post\_id, author, timestamp, text)}. The fallback is required because a small subset of exported records lacked comment identifiers.

Comment hierarchy (top-level vs.\ nested) is not available in these CSV exports, so all comments are treated at the thread level. This does not affect our primary timing estimand, which depends only on the earliest timestamp of any verification cue within a thread.

After preprocessing and de-duplication, the retained dataset used for cue matching and timing consisted of:
\begin{itemize}
  \item \textbf{r/moltbook}: 448 discussion threads (posts), 930 retained comments;
  \item \textbf{r/openclaw}: 417 discussion threads (posts), 938 retained comments.
\end{itemize}

All timestamps were normalized to UTC and represented as Unix epoch seconds. Engagement metadata (e.g., post score) reflects the value observed in the frozen API snapshot at the end of collection; because scores can evolve over time, we treat these quantities as snapshot measurements rather than stable properties. The dataset was frozen prior to analysis to ensure reproducibility, and no content outside the predefined observation window was included.

\subsection{Thread Construction}

Our unit of analysis is the discussion thread. Formally, we define a thread $T_i$ as a root post together with all retained comments associated with that post:
\[
T_i = \{p_i, c_{i1}, c_{i2}, \dots, c_{in_i}\},
\]
where $p_i$ denotes the original post and $c_{ij}$ denotes the $j$-th retained comment in the thread. Each post and comment is associated with a creation timestamp.

For each thread, we record the post creation time $t_i^{(0)}$, the post score $s_i$ (a platform-visible engagement signal), and the number of retained comments after de-duplication. Threads with zero retained comments are retained for prevalence analyses (e.g., whether a thread contains any verification cue). For timing analyses, such threads only contribute as right-censored observations in the censored-lag view.

\subsection{Verification Cues}

We operationalize verification using a conservative, rule-based lexicon that captures explicit requests for substantiation and evidence-providing behavior. The lexicon includes word-boundary patterns and phrase-level regular expressions such as:
\emph{source(s)}, \emph{citation(s)}, \emph{evidence}, \emph{proof}, \emph{link}, explicit URLs (\texttt{https?://}), and phrases indicating checking or attribution (e.g., \emph{I checked}, \emph{according to}, \emph{here is}, \emph{can you link}, \emph{verify}, \emph{verified}, \emph{where's the source/citation/proof/evidence}, \emph{show (me) the source/citation/proof/evidence}).
Matching is case-insensitive and uses regex word boundaries to reduce substring false positives.

A match is registered if any pattern appears anywhere in the comment body. Multiple matches within a comment are treated as a single match; multiple matches within a thread are collapsed to the earliest matched comment for timing.

For each comment $c_{ij}$, we define a binary indicator
\[
v_{ij} =
\begin{cases}
1, & \text{if } c_{ij} \text{ matches a verification cue}, \\
0, & \text{otherwise}.
\end{cases}
\]
A thread is considered to contain verification if $\sum_j v_{ij} \geq 1$.

\subsection{Verification Timing and Right-Censoring}

For threads that contain verification cues, we define the time-to-first-verification as the elapsed time between post creation and the first verification cue:
\[
\tau_i = \min_{j : v_{ij} = 1} (t_{ij} - t_i^{(0)}),
\]
where $t_{ij}$ is the timestamp of comment $c_{ij}$. We report timing in hours by converting seconds to hours.

If a matched comment has a missing timestamp, it contributes to verification \emph{presence} ($\sum_j v_{ij} \geq 1$) but is excluded from timing. When a matched comment appears with a timestamp earlier than the post timestamp, it is ignored for timing as a negative-lag anomaly; the earliest non-negative matched comment is used to define $\tau_i$. If no valid (timestamped, non-negative) matched comment exists, $\tau_i$ is undefined and the thread is excluded from conditional lag summaries, but remains part of prevalence tests.

For a supplementary right-censored view that incorporates threads where verification never appears, threads without a valid verification timestamp are treated as right-censored at the end of the observation window. For each subreddit $s$, let the dataset end time be the maximum timestamp observed in that subreddit’s frozen snapshot across all posts and retained comments:
\[
t^{(\mathrm{end})}_s = \max\Big(\{t_i^{(0)}\}_i \cup \{t_{ij}\}_{i,j}\Big).
\]
We define the censored time variable as
\[
\tilde{\tau}_i = \min(\tau_i,\, t^{(\mathrm{end})}_s - t_i^{(0)}),
\]
with censoring indicator
\[
\delta_i =
\begin{cases}
1, & \text{if } T_i \text{ contains a verification cue with a valid timestamp}, \\
0, & \text{otherwise (right-censored)}.
\end{cases}
\]
This corresponds to the computation of \texttt{censored\_lag\_hours} in the analysis code. Because this censored estimand includes threads without verification, it summarizes a different quantity than conditional lag medians, which are computed only among $\delta_i=1$ threads.

\subsection{Social Proof Grouping}

Platform-visible social proof is operationalized using post score. Within each subreddit, we compute the empirical score distribution under the frozen snapshot and define the high-visibility group as threads whose score lies at or above the 75th percentile:
\[
G_i =
\begin{cases}
\text{High}, & \text{if } s_i \geq Q_{0.75}(s), \\
\text{Low}, & \text{otherwise},
\end{cases}
\]
where $Q_{0.75}(s)$ is computed using linear interpolation on sorted values with position $\text{pos}=(n-1)q$. Under the frozen snapshot used for analysis, the $Q_{0.75}$ thresholds were 5.25 for \texttt{r/moltbook} and 3.00 for \texttt{r/openclaw}.

\subsection{Statistical Analysis}

We compare verification prevalence and verification timing across visibility groups using non-parametric statistics. For prevalence (whether a thread ever contains a verification cue), we use Fisher’s exact test and report odds ratios. For numerical stability, odds ratios are computed using the Haldane--Anscombe correction (+0.5 to each cell).

For timing, we focus on the distribution of time-to-first-verification conditional on a valid verification timestamp (i.e., $\delta_i=1$). We report group medians and IQRs and use permutation tests on the median difference. Specifically, we compute the observed median difference between groups and randomly permute group labels across threads within each subreddit for $N_{\text{perm}} = 5000$ iterations to obtain a two-sided $p$-value:
\[
p = \frac{\#\{|\Delta_{\text{perm}}| \ge |\Delta_{\text{obs}}|\} + 1}{N_{\text{perm}} + 1}.
\]
Permutation tests are deterministic given a fixed seed. Effect sizes are reported using Cliff’s delta, computed by pairwise comparisons of all observations across groups.

We do not fit regression models; all analyses are descriptive and focus on robust co-variation patterns rather than causal estimation.

\subsection{Audit and Reproducibility}

To assess lexicon precision, we manually audited matched comments. For the verification lexicon, we randomly sampled up to 25 matched comments per subreddit (using a fixed seed) and coded whether each match represented an unambiguous evidence-seeking or evidence-providing move (e.g., requesting a link/source, providing a link as substantiation). When fewer than 25 matches were available, we audited all matches; precision is reported with the labeled sample size.

We also audited additional cue families used in exploratory analyses (e.g., correction cues). When a cue family produced very few matches in a subreddit during the observation window, we audited all available matches up to 25. All analyses are deterministic and reproducible; no machine learning models are trained, and all parameters are explicitly specified.

\subsection{Ethical Considerations}

All data analyzed in this study were publicly available Reddit posts and comments. We report only aggregate statistics at the thread level and do not attempt to identify, track, or profile individual users. Our analyses focus on the prevalence and timing of verification cues rather than on evaluating the truthfulness of specific users’ claims.

\section{Results}

We report results in the order of figures. Throughout this section, we emphasize descriptive patterns and distributional differences rather than causal effects, and interpret findings in terms of verification timing and discussion structure rather than individual beliefs.

\begin{figure*}[t]
  \centering
  \includegraphics[width=\textwidth]{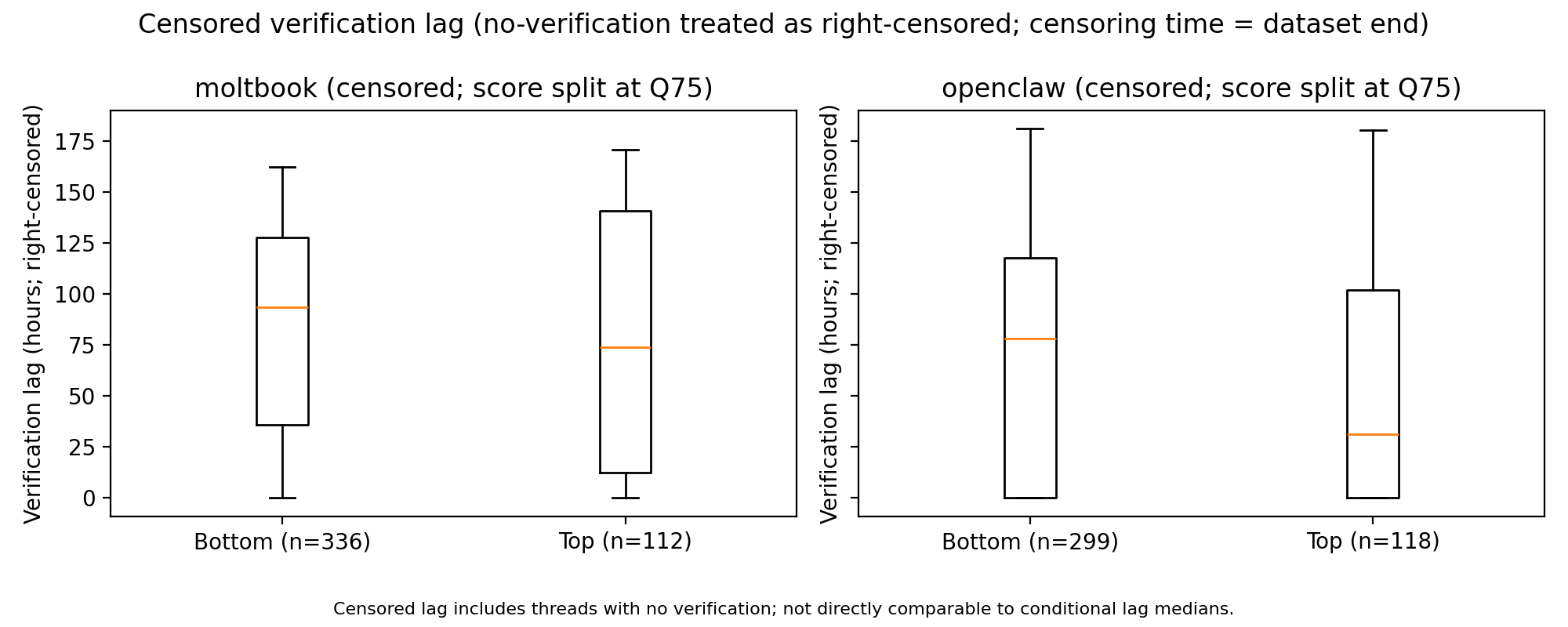}
  \caption{\textbf{Time-to-first-verification by platform visibility.} The distribution of time from post creation to the first verification cue (e.g., source, evidence, link) for each subreddit. Threads without any verification cue during the observation window are treated as right-censored at dataset end. High-visibility threads correspond to the top quartile of post score within each subreddit.}
  \label{fig:lag_censored}
\end{figure*}

\subsection{Delayed verification in high-visibility threads (Figure~\ref{fig:lag_censored})}

Figure~\ref{fig:lag_censored} shows the distribution of time-to-first-verification for discussion threads. Threads without any verification cue during the observation window are treated as right-censored at dataset end, so the figure jointly represents (a) delayed verification among threads that eventually contain verification and (b) complete absence of verification within the observation window.

Across both subreddits, high-visibility threads (top quartile by post score within each community) differ systematically from lower-visibility threads in verification timing. In \texttt{r/moltbook}, high-visibility threads exhibit markedly longer lags and a larger share of long-delay cases, producing an overall right-shifted distribution relative to low visibility. This separation is also visible when summarizing early verification coverage: among threads that contain verification cues, only 27\% of high-visibility \texttt{r/moltbook} cases show verification within 1 hour (ECDF$_{\leq1h}=0.27$), compared to 72\% in the low-visibility group; similarly, ECDF$_{\leq6h}$ is 0.57 (high) vs.\ 0.87 (low), and ECDF$_{\leq24h}$ is 0.81 (high) vs.\ 0.93 (low). These ECDF points make the same qualitative claim as the boxplots: when visibility is high, verification is less likely to happen early, and more likely to be delayed into later phases of the thread lifecycle.

In \texttt{r/openclaw}, many threads receive verification extremely quickly, resulting in a heavy mass at near-zero lag across both visibility groups. Even in this operational community, high visibility is associated with a longer-tailed delay distribution (Table~\ref{tab:verification_summary}). This is consistent with the broader pattern that high engagement can coincide with deferred evidence-seeking rather than earlier verification.

Importantly, these results do not imply that highly visible threads are less accurate, nor that participants are categorically less skeptical. Instead, they indicate a shift in the \emph{temporal organization} of evidence-seeking: when platform-visible engagement is high, verification is more likely to be deferred, allowing early narratives to circulate before substantiation is introduced.

\subsection{High-visibility threads attract verification more often, but later (conditional on occurring)}

Figure~\ref{fig:lag_censored} characterizes when verification appears (including right-censored threads). Here we separately quantify (1) whether verification occurs at all and (2) how late it arrives, conditional on occurring. Table~\ref{tab:verification_summary} summarizes these comparisons for both subreddits.

\begin{table}[t]
\centering
\caption{\textbf{Verification occurrence and delay by visibility group (score $\geq Q_{0.75}$).}
Occurrence uses Fisher's exact test with odds ratio (OR). Delay is conditional on verification occurring; we report median (IQR) hours, a permutation test on median difference, and Cliff's $\delta$.}
\label{tab:verification_summary}
\resizebox{\columnwidth}{!}{%
\begin{tabular}{lcc}
\toprule
 & \texttt{r/moltbook} & \texttt{r/openclaw} \\
\midrule
High-visibility verification rate & 33.0\% (37/112) & 54.2\% (64/118) \\
Low-visibility verification rate  & 16.1\% (54/336) & 35.1\% (105/299) \\
Fisher $p$; OR                   & $2.18\times10^{-4}$; 2.57 & $3.96\times10^{-4}$; 2.18 \\
\midrule
Lag (High), median (IQR), h      & 4.21 (0.93--16.81) & 0.00 (0.00--3.98) \\
Lag (Low), median (IQR), h       & 0.34 (0.19--1.28)  & 0.00 (0.00--0.09) \\
Permutation $p$; Cliff's $\delta$ & $4.00\times10^{-4}$; 0.47 & 0.005; 0.29 \\
\bottomrule
\end{tabular}
}%
\end{table}

Across both communities, high-visibility threads are more likely to contain at least one verification cue. In \texttt{r/moltbook}, 33.0\% of high-visibility threads (37/112) contain verification cues compared to 16.1\% of low-visibility threads (54/336), yielding Fisher's exact test $p=2.18\times10^{-4}$ and odds ratio (OR) $=2.57$. In \texttt{r/openclaw}, 54.2\% of high-visibility threads (64/118) contain verification cues compared to 35.1\% (105/299), with Fisher $p=3.96\times10^{-4}$ and OR $=2.18$. Interpreting OR in this setting, high visibility roughly doubles the odds that a thread will receive at least one explicit evidence-seeking move during the observation window.

However, increased likelihood of verification does not imply earlier verification. Conditional on verification occurring, \texttt{r/moltbook} shows a substantial timing shift: high-visibility threads have a median lag of 4.21h (IQR 0.93--16.81), whereas low-visibility threads have a median lag of 0.34h (IQR 0.19--1.28). A permutation test on median lag yields $p=4.00\times10^{-4}$, and Cliff's $\delta=0.47$ indicates that the high-visibility distribution is overall right-shifted relative to low visibility. Put plainly, when verification happens in \texttt{r/moltbook}, it tends to arrive substantially later in high-visibility discussions.

In \texttt{r/openclaw}, both groups have a median conditional lag of 0.00h because many threads contain immediate verification cues. Nonetheless, the dispersion differs sharply: the high-visibility group has a much wider IQR (0.00--3.98h) than the low-visibility group (0.00--0.09h). This difference is statistically detectable (permutation $p=0.005$; Cliff's $\delta=0.29$) and indicates a heavier right tail in high-visibility threads. This is a case where the median alone is not informative: while many high-visibility threads are verified immediately, a non-trivial subset experiences substantially delayed verification, producing a longer-tailed delay profile than the low-visibility group.

Taken together, these results support a consistent qualitative pattern across both communities: high-visibility threads attract verification more often, but verification tends to occur later (and sometimes not at all within the window), meaning evidence-seeking is present yet temporally displaced.

\subsection{Verification timing differs by discourse context}

While the visibility--verification relationship appears in both subreddits, its magnitude and shape vary by discourse context. In \texttt{r/openclaw}, where operational norms make evidence-seeking and troubleshooting more common, verification frequently appears immediately, compressing both groups toward near-zero lags and shifting the visibility signal into the distribution tail (Table~\ref{tab:verification_summary}). In contrast, \texttt{r/moltbook} exhibits both larger differences in conditional central tendency and lower early-verification coverage, consistent with a setting in which exploratory and interpretive discussion can proceed for longer periods before participants introduce explicit substantiation requests.

\begin{figure*}[t]
  \centering
  \includegraphics[width=\textwidth]{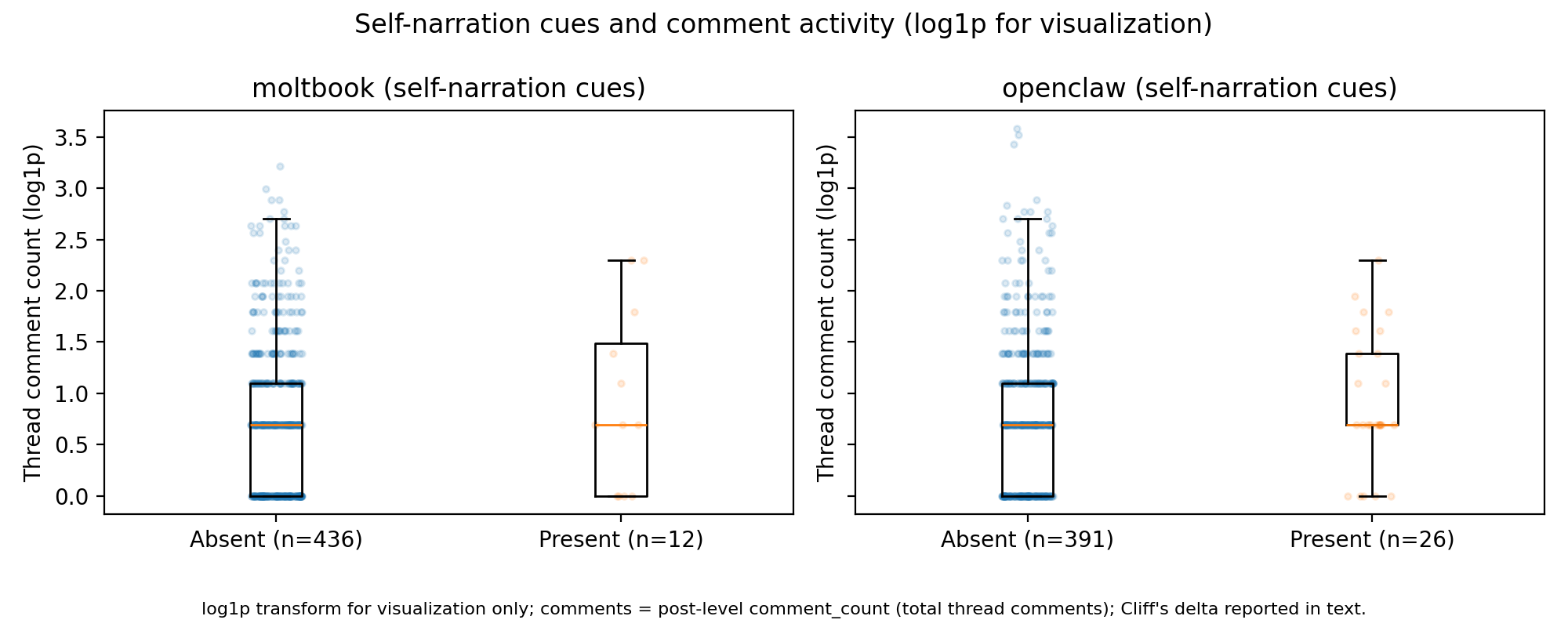}
  \caption{\textbf{Comment activity and self-narration cues.} Thread-level comment counts (log1p) for posts with and without self-narration cues. Self-narration cues capture affective or personal language co-occurring with references to agentic AI. Due to the small number of cue-present threads, results are descriptive.}
  \label{fig:self_narration}
\end{figure*}

\subsection{Self-narration cues and comment activity (Figure~\ref{fig:self_narration})}

Figure~\ref{fig:self_narration} presents an exploratory comparison of thread comment activity for posts containing self-narration cues versus those that do not. Self-narration cues are rare in both subreddits, and the small number of cue-present threads limits inferential interpretation.

Across both communities, cue-present threads do not exhibit systematically higher comment counts. We include this analysis to probe a simple alternative explanation: that the observed verification delay in high-visibility threads is primarily driven by emotive or story-like discourse that independently increases engagement. Instead, the observed pattern in Figure~\ref{fig:lag_censored} aligns more directly with platform-visible social proof and verification timing, suggesting that the temporal displacement of verification is not reducible to narrative style differences alone.

In summary, platform-visible engagement is associated with a shift in the temporal structure of evidence-seeking in public discussions of agentic AI. Highly visible threads are more likely to attract verification eventually, but verification tends to occur later, often after early narratives have already gained traction. Rather than directly measuring individual trust, these results characterize a pre-interaction trust environment in which credibility can be inferred from engagement before evidence is introduced.

\section{Discussion}

This work examines how public discussions of agentic AI unfold under platform-visible engagement signals, and how the timing of verification shapes the informational conditions under which trust is formed. Rather than evaluating the accuracy of claims or measuring individual trust directly, we focus on verification timing as a structural property of discourse. This perspective connects our findings to broader theories of social proof, collective cognition, and trust calibration in novel sociotechnical systems \cite{Lee2004Trust, Muchnik2013Social, Avram2020Exposure}.

We focus this discussion on two implications that follow directly from our results: (1) platform-visible engagement can bias the pre-interaction trust environment by shifting users toward credibility-by-visibility, and (2) late-arriving evidence can be structurally disadvantaged because timing shapes whether and how corrections update beliefs.

\subsection{From evidence-seeking to credibility-by-visibility}

A central finding is that high-visibility discussions exhibit delayed or absent verification early in the thread lifecycle. This matters because, in pre-interaction settings, users often lack direct experience with the systems being discussed and cannot easily evaluate claims through personal testing. Under these conditions, platform-visible engagement becomes a salient proxy for credibility \cite{Avram2020Exposure}. Social influence bias further implies that visible popularity can shift judgments and attention even when underlying quality is held constant \cite{Muchnik2013Social}.

Our results suggest that this proxy is not only associated with \emph{what} people attend to, but with \emph{when} they decide evidence is necessary. If a thread appears ``already validated'' by attention, participants may reasonably infer that verification has been performed by others, reducing the perceived need for immediate evidence-seeking. This is a diffusion of epistemic responsibility: verification is implicitly delegated to the crowd. In this sense, engagement signals can shape a pre-interaction trust ecology that favors credibility-by-visibility rather than credibility-by-evidence, which is precisely the type of miscalibration risk highlighted in trust-in-automation research \cite{Lee2004Trust}.

This mechanism is consistent with a narrative lock-in interpretation. Early claims can structure attention and interpretation before substantiation is visible, and later contributions may be forced to respond within the frame set by the early narrative. Importantly, narrative lock-in here does not assume that participants are naive or unskeptical. Instead, skepticism becomes temporally displaced. Verification can still occur, but it tends to arrive later, after the thread has already accumulated attention and conversational momentum.

Two clarifications are important for keeping this interpretation within the limits of what our data support. First, we do not claim that high visibility causes reduced scrutiny; we show that high visibility is associated with a different temporal organization of scrutiny. Second, our operationalization captures \emph{explicit} evidence-seeking (e.g., requests for sources, links, or ``I checked'' statements). High-visibility threads may also contain implicit skepticism, technical disagreement, or sarcasm that does not trigger our lexicon. In that sense, the pattern we report should be interpreted as a conservative lower bound on delayed \emph{explicit} verification rather than a complete accounting of all critical engagement.

\subsection{Why timing matters: late evidence may update less}

A second implication follows from the combination of delayed verification and established timing effects in correction and misinformation research. Work on the continued influence effect shows that misinformation can persist even after corrections appear \cite{Lewandowsky2012Misinformation}. Empirical evidence also indicates that timing matters: later corrections can be less effective than earlier interventions \cite{Brashier2021Timing}, while proactive accuracy prompts can improve truth discernment \cite{Pennycook2021Shifting}. These findings imply that, even when verification eventually arrives, its capacity to revise beliefs and interpretations can be constrained by what has already become salient and coherent in the thread.

Our contribution is to show how platform-visible engagement can make ``late evidence'' more likely in the first place. When verification is structurally delayed in high-visibility threads, the discourse environment increases the chance that participants encounter a narrative \emph{before} they encounter substantiation. Even if evidence appears later, its informational role is different. It must compete with an already-established conversational trajectory, and it may be treated as a post-hoc addendum rather than a foundational constraint. In practical terms, delayed verification shifts evidence-seeking from early sorting to late repair, which prior work suggests is often disadvantaged \cite{Lewandowsky2012Misinformation, Brashier2021Timing, Pennycook2021Shifting}.

This framing also clarifies why we do not need to claim a direct causal effect on individual trust to make the argument meaningful. Even without measuring trust outcomes, shifting when and whether evidence is requested changes the informational environment in which trust calibration is likely to occur \cite{Lee2004Trust}. In other words, verification timing is a property of the environment that conditions subsequent judgments, rather than a direct measurement of those judgments.

\subsection{Implications for agentic AI and platform design}

Our findings suggest actionable directions for both agentic AI developers and platform designers.

For agentic AI developers, public discourse should be treated as part of the system’s trust surface. Expectations formed through online discussion can influence adoption, reliance, and oversight behaviors. Supporting early verification cues in public-facing documentation and demos, such as clear provenance, reproducible claims, and links to independent benchmarks, may reduce reliance on engagement as a stand-in for credibility and support more appropriate reliance \cite{Lee2004Trust}. A practical implication is that ``capability claims'' should be packaged with verifiable artifacts that are easy to cite in discussion threads (e.g., stable links, minimal reproduction steps, and explicit caveats), because the presence of shareable evidence can lower the friction of early verification.

For platform designers, the results highlight an epistemic risk of engagement-driven ranking: visibility can outpace verification. Design friction offers plausible, lightweight interventions that preserve participation while encouraging reflective evaluation \cite{Distler2020, Ribeiro2023, Tseng2025}. To make these implications concrete in the Reddit-like setting we study, interventions can be specified as small interface changes with explicit triggers and user actions, for example: (1) a ``source prompt'' that appears when a post about agentic AI capability exceeds a visibility threshold, asking the author to optionally add a link or label the claim as speculative; (2) a verification-status badge (e.g., ``evidence requested'' vs.\ ``evidence linked'') that becomes salient in high-traction threads; and (3) surfacing unanswered verification requests near the top of the thread so that late-arriving readers see epistemic gaps before they see the most upvoted narrative. Importantly, the goal is not to suppress discussion, but to rebalance incentives so that visibility is less likely to function as a substitute for evidence.

\subsection{Limitations and future work}

This study has several limitations. We analyze discussions from two Reddit communities, which may not generalize to other platforms or cultural contexts. Our lexical approach prioritizes precision over recall and may miss subtler forms of verification. In particular, it can undercount (1) implicit skepticism that is expressed without explicit ``source/evidence/link'' language, (2) sarcasm or rhetorical doubt, and (3) technical verification embedded in implementation talk that does not use our cue terms. Because our signal is intentionally conservative, our findings should be interpreted as a lower bound on delays in \emph{explicit} evidence-seeking. Additionally, our analyses are descriptive; we do not estimate causal effects of visibility on verification timing.

A clear next step is a participant study that tests whether the timing patterns we observed translate into differences in trust calibration, belief updating, and memory of the discussion. Our results motivate two experimental manipulations that directly operationalize the mechanisms implied by the observational findings: (1) \textit{platform-visible social proof} (high vs.\ low apparent popularity), and (2) \textit{verification timing} (early vs.\ late vs.\ absent verification cues). A minimal $2 \times 3$ design can present participants with realistic, Reddit-like threads about agentic AI capability claims where the claim content is held constant but the thread metadata and evidence timing vary.

Primary outcomes should map to the conceptual claims of this paper. First, \textit{trust calibration and reliance intentions}: perceived credibility of the claim, willingness to adopt or recommend the system, and perceived need for personal verification \cite{Lee2004Trust}. Second, \textit{belief updating}: participants provide an initial judgment after seeing the post and early comments, then re-rate after the verification cue appears (or does not), allowing estimation of how much late evidence shifts beliefs relative to early evidence. Third, \textit{episodic summary and recall}: immediately and after a delay, participants summarize ``what happened'' in the thread and answer questions about what evidence was provided, enabling tests of whether high-visibility narratives dominate the remembered gist when evidence arrives late. These outcomes directly connect the timing mechanism to the literature on continued influence and correction timing \cite{Lewandowsky2012Misinformation, Brashier2021Timing, Pennycook2021Shifting}.

To connect the experiment tightly to our field measures, the stimuli can be parameterized using the empirical lag distributions from our data (e.g., selecting early/late verification timings that correspond to typical lower-visibility vs.\ high-visibility lags). This allows direct translation between the observational timing gaps and the experimental manipulation magnitudes. If the hypothesized mechanism holds, we would expect high social proof to reduce early evidence-seeking intentions and to attenuate belief updating when verification arrives late, consistent with the idea that late evidence is structurally disadvantaged \cite{Lewandowsky2012Misinformation, Brashier2021Timing, Pennycook2021Shifting}.

Beyond vignette studies, a complementary next step is an intervention study that evaluates whether lightweight friction or provenance prompts can shift verification earlier in high-traction threads. Such interventions can be assessed using the same time-to-first-verification and right-censoring framework used here, enabling direct comparability between observational measurement and intervention effects \cite{Distler2020, Ribeiro2023, Tseng2025}. Together, these follow-ups would clarify whether the timing patterns we document are merely descriptive correlations or whether they causally shape trust formation and correction effectiveness in public discourse about agentic AI.

\section{Conclusion}

This paper examined how public discussions of agentic AI evolve under platform-visible engagement signals, and how verification timing shapes the informational conditions in which trust is formed. Using Reddit discussions as an in-the-wild setting, we showed that highly visible threads systematically differ in evidence-seeking behavior: verification is more likely to occur eventually, but tends to be delayed or absent during the early stages of discussion. As a result, early narratives can gain traction and structure collective understanding before substantiation is introduced.

We interpret this pattern as consistent with collective cognitive bias under conditions of novelty and epistemic uncertainty. In such contexts, platform-visible engagement can function as a salient proxy for credibility, encouraging participants to delegate verification to the crowd rather than initiating it themselves \cite{Muchnik2013Social, Avram2020Exposure}. Importantly, this does not imply reduced skepticism or irrationality; rather, it reflects a shift in verification timing: evidence-seeking is more likely to occur after discussion has already accumulated, rather than near the start of the thread.

By focusing on verification timing rather than content accuracy or individual attitudes, this work contributes a complementary perspective to existing research on misinformation and social proof \cite{Lewandowsky2012Misinformation, Pennycook2021Shifting}. 
It also extends trust-in-AI and trust calibration work by characterizing a pre-interaction information environment where credibility is inferred from visibility before evidence appears \cite{Lee2004Trust}. Our findings highlight that trust in agentic AI is shaped not only by system behavior or user experience, but also by the structure of public discourse that precedes direct interaction. Designing platforms and AI systems that support timely and visible verification may help preserve epistemic agency and promote more calibrated trust as autonomous AI becomes increasingly integrated into everyday life.

\bibliographystyle{ACM-Reference-Format}
\bibliography{references}

\end{document}